\documentstyle[epsfig,12pt]{article}

\begin{document}
\title{Interlinkages and structural changes in cross-border liabilities: a
network approach}
\begin{small}
\author{Alessandro Spelta* and Tanya Ara\'{u}jo ** \\
(*) University of Pavia, Italy \\
\emph{alessandro.spelta@universitapavia.it}\\
 ISEG - Technical University of Lisbon (TULisbon) and\\
Research Unit on Complexity in Economics (UECE), Portugal\\
\emph{tanya@iseg.utl.pt}} \end{small}
\date{}
\maketitle

\begin{abstract}
\begin{small}
We study the international interbank market through a geometrical and a
topological analysis of empirical data. The geometrical analysis of the time
series of cross-country liabilities shows that the systematic information of
the interbank international market is contained in a space of small
dimension, from which a topological characterization could be conveniently
carried out. Weighted and complete networks of financial linkages across
countries are developed, for which continuous clustering, degree centrality
and closeness centrality are computed. The behavior of these topological
coefficients reveals an important modification acting in the financial
linkages in the period 1997-2011. Here we show that, besides the generalized
clustering increase, there is a persistent increment in the degree of
connectivity and in the closeness centrality of some countries. These
countries seem to correspond to critical locations where tax policies might
provide opportunities to shift debts. Such critical locations highlight the
role that specific countries play in the network structure and helps to
situates the turbulent period that has been characterizing the global
financial system since the Summer 2007 as the counterpart of a larger
structural change going on for a more than one decade.
\end{small}
\end{abstract}

\begin{small}
\textbf{Keywords:} Cross-border exposures, interbank networks, financial
linkages, debt shifting \\

\emph{Corresponding author: tanya@iseg.utl.pt}
\end{small}

\section{Introduction}

Globalization of economies leads to an ever-increasing interdependence of
countries. The late-2000s financial crisis - considered by many economists
to be the worst financial crisis since the Great Depression - resulted in
the collapse of large financial institutions, the bailout of banks by
national governments and downturns in stock markets around the world. Such a
large set of outcomes has stressed the need of understanding financial
systems as networks of countries where cross-border financial linkages play
the fundamental role.

While some authors have investigated the role globalization plays in shaping
the spread of financial crisis \cite{Soros},\cite{Akerlof}, studies on the
consequences of financial crises to the international banking system are
less prominent. As we recently argued \cite{Spelta}, the adoption of an
evolving network approach is recommended not only because of the proper
emphasis on the financial interdependencies but also due to the possibility
of describing how the structure of these interdependencies evolves in time.
In so doing, we are able to address the role that an existing network
structure plays in the spread of shocks and conversely, the effectiveness of
stress events and their impact on the structure of the network of
cross-border interdependencies.

During the last years, several authors have approached financial systems
through a network approach. Some papers have favored the study of
interdependencies between credit banks \cite{Kubelec}, or focused on the
analysis of shocks storming the financial systems of several countries \cite%
{Helpman}. The topological properties of some national interbank markets
have been studied by Soramaky and co-authors \cite{Soromaki}, who analyzed
the network topology of the interbank payments transferred between
commercial banks by the Fedwire Funds Services. Another example is the work
of Fuijwara \cite{Fujiwara} exploring the credit relationships that exist
between commercial banks and large companies in Japan.

Empirical studies have also been carried out on some European national
interbank markets (\cite{Masi},\cite{Boss}) throughout the analysis of the
topological properties of the networks of Italian and Austrian banks.

Using the Bank of international settlements (BIS) data set, some papers have
addressed the evolution of networks of bank transfers (\cite{Boss}, \cite%
{Garratt}, \cite{Fujiwara}, \cite{Kubelec}, \cite{McGuire} ). It is evident
from the work of McGuire \cite{McGuire} that the international banking
system has become an important conduit for the transfer of capital across
countries. The work of Minoiu \cite{Minoiu} also explored the proprieties of
the global banking networks; for several networked systems, he found
evidence of important structural changes follow on the occurrence of stress
events.

Nevertheless, yet few papers have addressed the impact of the recent
financial crises in the international banking system from a topological
point of view.

Topological coefficients have been the object of growing attention ever
since some network regimes were identified as the underlying structures of
important phenomena found in many different fields.

Stock market crises were characterized in the past through a geometrical and
a topological analysis of the time series of stock returns. It implied that
the most of the systematic information of the financial markets was
contained in a space of small dimension (\cite{Vilela},\cite{Araujo},\cite%
{Araujo1},\cite{Araujo2}) from which both geometrical and topological
characterizations could be carried out.

In the present paper, we explore an equivalent low-dimensional space to find
out the topological signature of networks of financial linkages across
countries. The networks are built from time series of interbank liabilities
conducted through the international banking system along the last 28 years.

From the geometrical setting, the behavior of the low-dimensional spaces
highlights an important modification acting in the financial linkages in the
period 1997-2011, and situates the turbulent period that has been
characterizing the global financial system since the Summer 2007 as the
counterpart of a larger structural change going on for more than one decade.

From the topological perspective, our results reveal an important increase
in the coefficient of clustering since 2006. They also show that the most
connected countries (largest degrees) are typically not the ones with the
largest closeness centrality. This is observed in different time periods,
allowing to identify some structural patterns that emerge from the evolution
of the network of country debts.

Here, we identify the chronological chain of countries with the highest
closeness centrality and hypothesize that their critical locations in the
network of countries are associated to low taxes policies, since they have
been considered inward recipients of debt shifting, due to a lower effective
taxation of foreign banks.

The paper is organized as follows: Section 2 briefly presents the set of
empirical data and the methodology we work with. Section 3 is targeted at
presenting the first results obtained from the application of a stochastic
geometry technique. The main contributions of the paper are presented in
Section 4, where a network approach is used to analyze the structural
changes in the cross-border liabilities across countries along the last 28
years. The paper ends with appropriate conclusions.

\section{Data and Methodology}

The Bank for International Settlements (BIS) locational banking statistics
(IBLR) - including international claims and liabilities of reporting banks
by country of residence - provides a plentiful data set of aggregate
cross-border exposures for a set of reporting and non-reporting countries
all over the world. These statistics were originally introduced in 1964 to
monitor the development of euro-currency markets, starting to be available in 1977
\footnote{The statistics prior to 1983 include just fifteen countries.}. The locational
reporting system collects quarterly data on the gross international
financial claims and liabilities of banks resident in a given country. The
main purpose of the statistics is to provide information on the role of
banks and financial centers in the intermediation of international capital
flows.

Since the BIS locational banking statistics capture the net flows of
financial capital between any two countries channeled through the banking
system \cite{McGuire} this data set is an appropriate source to the
empirical study of temporal patterns arising from financial linkages across
countries.

The locational statistics are intended to complement monetary and credit
aggregates, being consistent with both the national balances of payments and
the systems of national accounts. For a set of 57 reporting countries and
aggregate zones, we considered a subset of 24 countries (see Table 1), each of them represented by the
amounts of its liabilities \textit{vis-a-vis} the other reporting countries,
measured on a quarterly basis, from the last quarter of 1983 (Q4-1983) to
the third quarter of 2011 (Q3-2011).

\begin{center}
\begin{tabular}{|c|c|}
\hline
AT: Austria & IT: Italy \\ \hline
BS: Bahamas & JP: Japan \\ \hline
BH: Bahrain & LU: Luxemburg \\ \hline
BE: Belgium & NL: Netherlands \\ \hline
CA: Canada & AN: Netherlands Antilles \\ \hline
KY: Cayman Islands & NO: Norway \\ \hline
DK: Denmark & SG: Singapore \\ \hline
FI: Finland & ES: Spain \\ \hline
FR: France & SE: Sweden \\ \hline
DE: Germany & CH: Switzerland \\ \hline
HK: Hong Kong & GB: United Kingdom \\ \hline
IE: Ireland & US: United States \\ \hline
\end{tabular}

Table 1: Reporting Countries
\end{center}

\subsection{The Method}

Cross-correlation based distances, as applied in reference \cite{Mantegna}
to the study of stock market structures have been used in the analysis and
reconstruction of geometric spaces in many different fields (\cite{Vilela},
\cite{Araujo1}, \cite{Araujo2}). The quantity

\begin{equation}
d_{ij}=\sqrt{2\left( 1-C_{ij}\right) }  \label{I.1}
\end{equation}

where $C_{ij}$ is the correlation coefficient of two time series $%
\overrightarrow{s}(i)$ and $\overrightarrow{s}(j)$ computed along a given
time window
\begin{equation}
C_{ij}=\frac{\left\langle \overrightarrow{s}(i)\overrightarrow{s}%
(j)\right\rangle -\left\langle \overrightarrow{s}(i)\right\rangle
\left\langle \overrightarrow{s}(j)\right\rangle }{\sqrt{\left( \left\langle
\overrightarrow{s}^{2}(i)\right\rangle -\left\langle \overrightarrow{s}%
(i)\right\rangle ^{2}\right) \left( \left\langle \overrightarrow{s}%
^{2}(j)\right\rangle -\left\langle \overrightarrow{s}(j)\right\rangle
^{2}\right) }}  \label{I.2}
\end{equation}

has been shown \cite{Stanley} to satisfy all the metric axioms. Therefore,
it may be used to develop a geometrical analysis of the interbank market
structure. To this end, we applied the following technique.

\subsection{The Stochastic Geometry Technique}

Using the BIS quarterly based time series of liabilities of each country ($%
\overrightarrow{l}(i)$) \textit{vis-a-vis} the other reporting countries, we
define a normalized vector

\begin{equation}
\overrightarrow{\rho }(i)=\frac{\overrightarrow{l}(i)-\left\langle
\overrightarrow{l}(i)\right\rangle }{\sqrt{n\left( \left\langle
\overrightarrow{l}^{2}(i)\right\rangle -\left\langle \overrightarrow{l}%
(i)\right\rangle ^{2}\right) }}  \label{2.2}
\end{equation}

$n$ being the number of components (number of time labels) in the vectors $%
\overrightarrow{l}(i)$. With this vector one defines the \textit{distance}
between the countries $i$ and $j$ by the Euclidean distance of the
normalized vectors,

\begin{center}
\begin{equation}
d_{ij}=\sqrt{2\left( 1-C_{ij}\right) }=\left\Vert \overrightarrow{\rho }(i)-%
\overrightarrow{\rho }(j)\right\Vert
\end{equation}
\end{center}

where $C_{ij}$ is the correlation coefficient of the liabilities $%
\overrightarrow{l}(i)$ and $\overrightarrow{l}(j)$, respectively of
countries $i$ and $j$ computed along a time window of $n$ (quarterly)
observations.

Having computed the matrix of distances for the set of $N$ reporting
countries, one obtains coordinates in $R^{N-1}$ compatible with these
distances and the countries can now be represented by a set $\left\{
x_{i}\right\} $ of points in $R^{N-1}$. Then, the center of mass $%
\overrightarrow{R}$ and the center of mass coordinates $\overrightarrow{y}%
(k)=\overrightarrow{x}(k)-\overrightarrow{R}$ are calculated.

The covariance distance matrix

\begin{center}
\begin{equation}
T_{ij}=\Sigma _{k}\overrightarrow{y_{i}}(k)\overrightarrow{y_{j}}(k)
\label{I.3}
\end{equation}
\end{center}

is diagonalized to obtain the set of eigenvalues and normalized eigenvectors
\{$\lambda _{i},\overrightarrow{e}_{i}$\}. The eigenvectors $\overrightarrow{%
e}_{i}$ define the characteristic directions of this geometric space. Their
coordinates $z_{i}(k)$\ along these directions are obtained by projection
\begin{equation}
z_{i}(k)=\overrightarrow{y}(k)\bullet \overrightarrow{e}_{i}
\end{equation}

The same analysis is performed for random and time permuted data and the
relative behavior of the eigenvalues is compared.

The characteristic directions correspond to the eigenvalues $\lambda _{i}$
that are clearly different from those obtained from surrogate data. They
define a subspace $V_{d}$ of dimension $d$ which contains the systemic
information related the interbank market structure.

Having found the number of characteristic dimensions ($d$) we are able to
denote by $\overrightarrow{z}(k)^{(d)}$ the restriction of the $k-$country
to the subspace $V_{d}$ and by $d_{ij}^{(d)}$ the distances between
countries $i$ and $j$ restricted to this low-dimensional space.

\section{First Results}

Our study starts with the application of the stochastic geometry technique
to the time series of interbank liabilities. Each time series corresponds to
one reporting country and provides quarterly observations over 28 years,
from the last quarter of 1983 to the third quarter of 2011.

\subsection{The Dimensional Reduction}

Having carried out this analysis for the 24 reporting countries over the 28
years we concluded that the systematic information related to the interbank
market structure is contained in a reduced subspace of just three
dimensions, meaning that the number of characteristic dimensions of this
subspace is three ($d=3$).

Figure 1(a) shows the ordered eigenvalue distributions of actual ($\lambda
_{i}$) and random ($\lambda _{i}^{\prime }$) data. From the difference
between the decay of actual and random (time-permuted) eigenvalues we
conclude that the first three dimensions capture the structure of the
deterministic correlations and economic trends that are driving the
cross-border financial system, whereas the remainder of the space may be
considered as being generated by random fluctuations.

Eigenvalues from the 4th to the 23th have a quite similar behavior in both
actual and time-permuted cases, showing that for this cross-border system
the three largest dimensions define our empirically constructed variables.

\begin{figure}[htb]
\begin{center}
\includegraphics[width=0.45\linewidth]{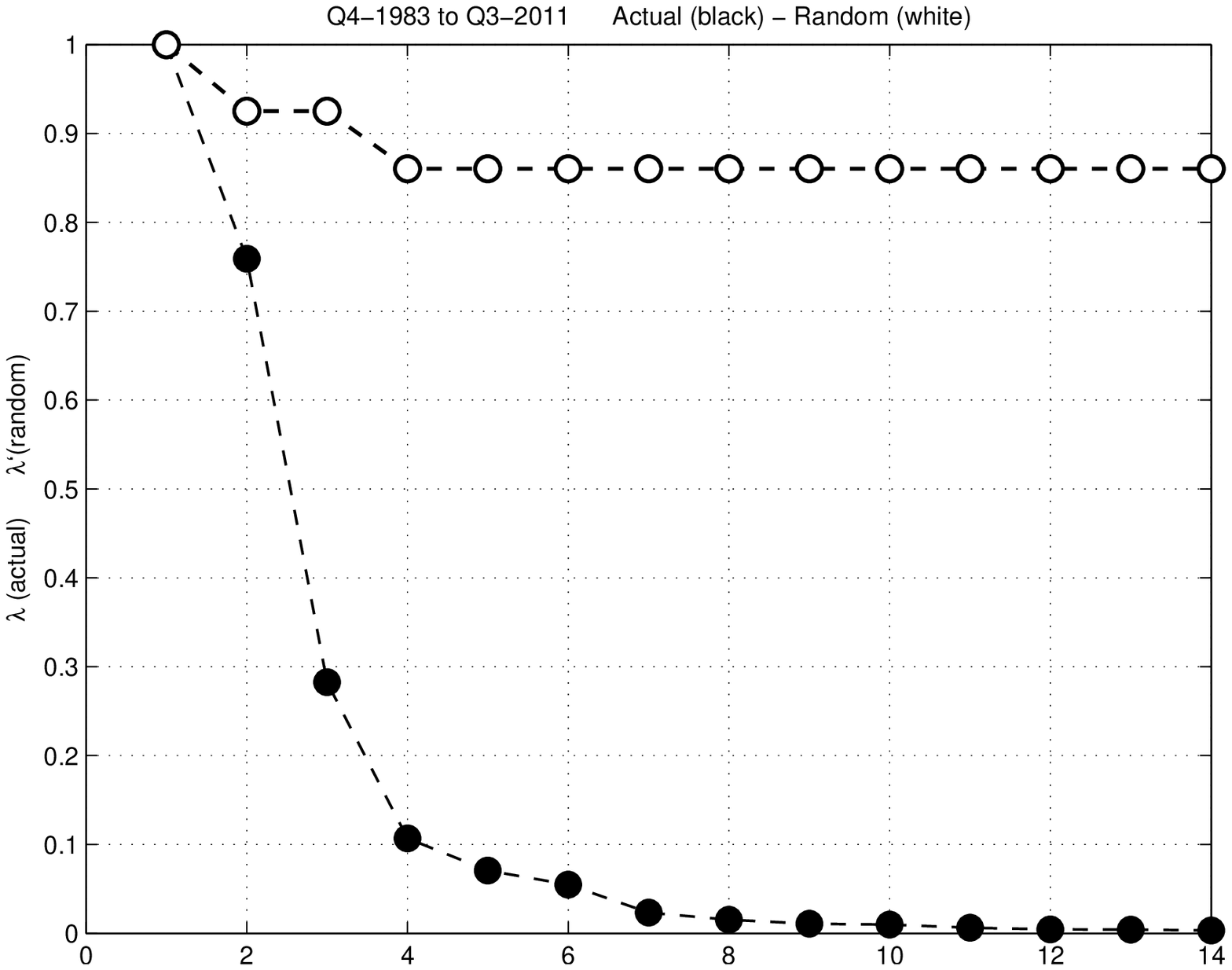}\includegraphics[width=0.45%
\linewidth]{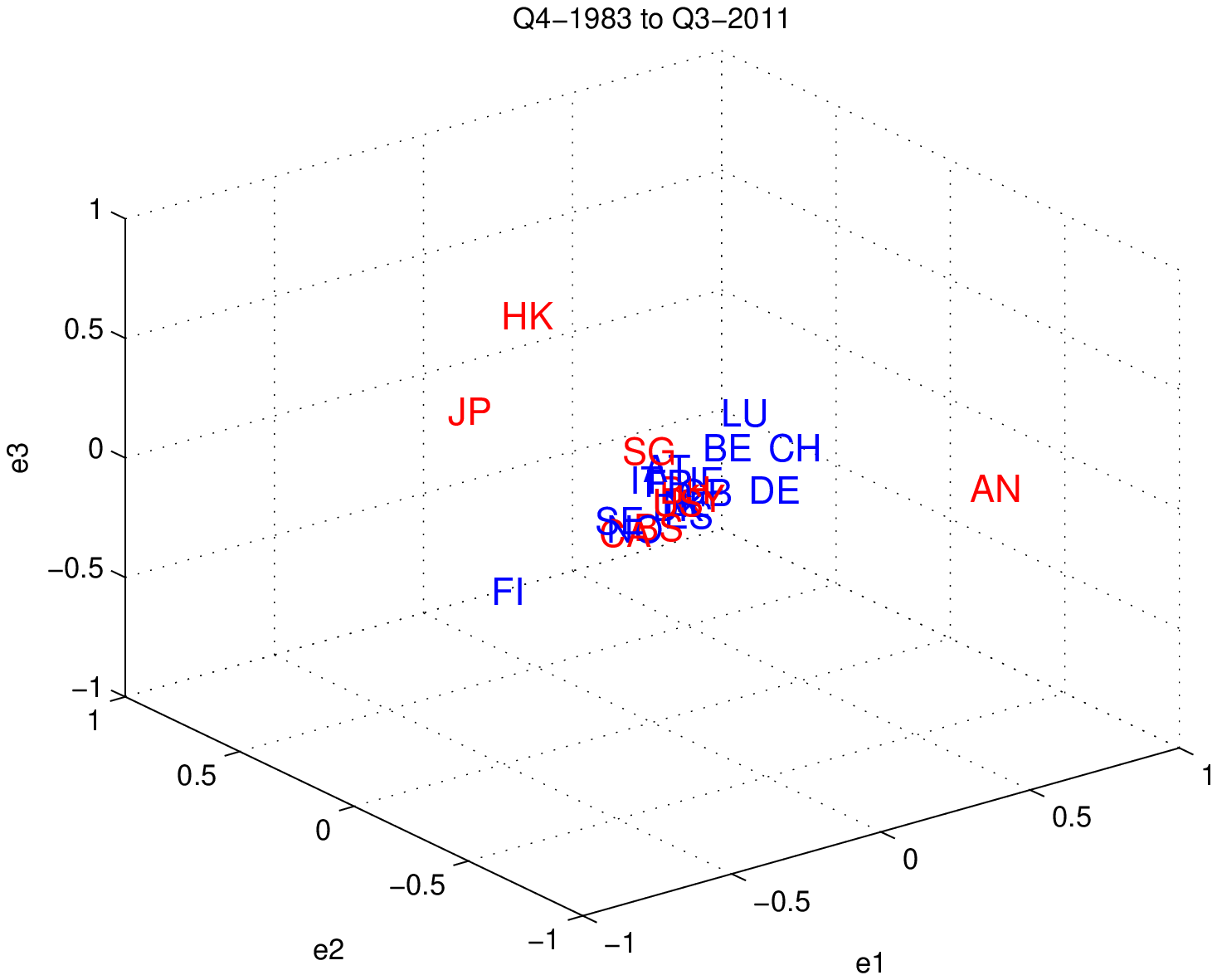}
\end{center}
\caption{ \emph{(a) }The ordered eigenvalue distributions (actual/black and
random/white) and \emph{(b)} the $\emph{3-}$dimensional space 1983-2011}
\end{figure}

Figure 1(b) shows the projections ($\overrightarrow{z}(k)^{(3)}$) of the 24
countries along the first three eigenvectors obtained from the entire
28-years actual data ($N=24$ and $n=110$ quarters), over the period
1983-2011. Different colors are used to identify European (blue) and
non-European (red) member states.

The \emph{3-}dimensional plot in Figure 1(b) shows that the most eccentric
positions in the reduced subspace are occupied by countries outside the
European Union (EU). Countries like Japan (JP), Hong Kong (HK) and the
Netherlands Antilles (AN) are far away from all other countries, whereas a
much greater centrality (and closeness) is characteristic of many EU
countries. Notice that to be close to each other in the \emph{3-}dimensional
space comes from positive and strong correlations between country debts.
Conversely, the countries that split apart from the others have debt
positions that are negatively (and strongly) correlated to the debt
positions of the other countries. In this example and when the whole 28
years are considered, Finland (FI), Japan (JP), the Netherlands Antilles
(AN) and Hong Kong (HK) are strongly and negatively correlated to the great
majority of reporting countries.

To have a qualitative idea concerning the structure of the characteristic
dimensions, we have divided the data in two chronologically successive
batches and performed the same operations. The subplots in Figure 2 show the
reduced subspaces associated to the three largest eigenvalues obtained over
the 14-years periods 1983-1997 and 1997-2011.

Apart from statistical fluctuations, the reconstructed sub-spaces in Figure
2 show a reasonable degree of stability. The ordering of the largest
eigenvalues changes in time although the overall distribution remains
approximately the same. These ordering change may have an economic meaning
and be related to the relative importance and stability of groups of
countries in different periods.

\begin{figure}[htb]
\begin{center}
\includegraphics[width=0.8\linewidth]{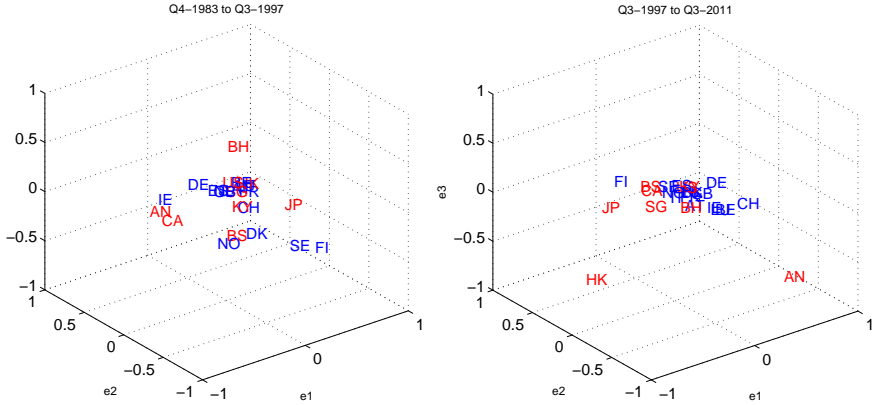}
\end{center}
\caption{The \emph{3-}dimensional spaces: 1983-1997 (left) and 1997-2011
(right)}
\end{figure}

Although stable, the geometrical spaces in Figure 2 show that there are some
important differences in the interbank market of those different time
periods. The main difference seems to rely on the \emph{clustering trend }%
that characterizes the 1997-2011 geometrical space, where the distances
between countries seem to be shortened; except for the few examples of the
Netherlands Antilles (AN) and Hong Kong (HK) that like in the subplot
presented in Figure 1(b) remain far from almost all other countries.

\begin{figure}[htb]
\begin{center}
\includegraphics[width=0.7\linewidth]{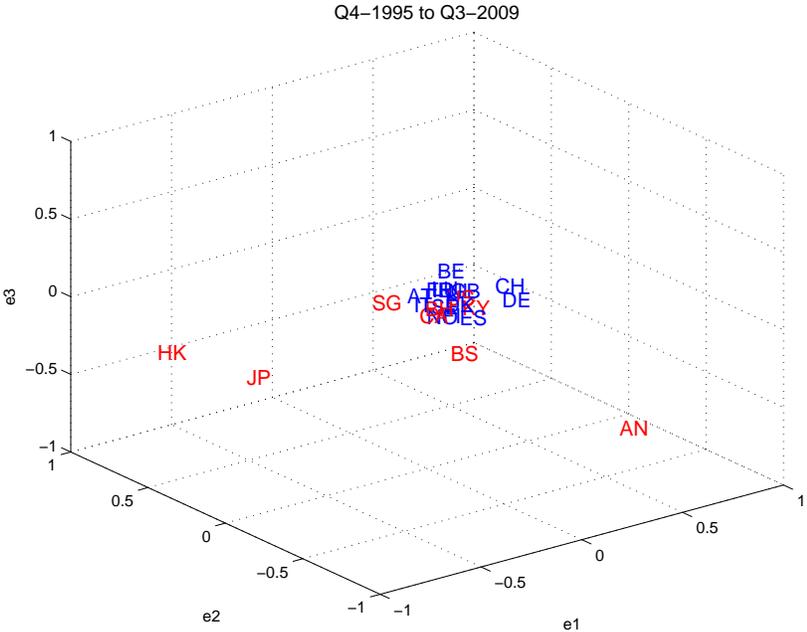}
\end{center}
\caption{The \emph{3-}dimensional space 1995-2009}
\end{figure}

To better investigate the shortening of the distances between countries that
may occur in different historical periods, the plot in Figure 3 provides the
geometrical space of another 14-years period, from 1995 to 2009. Again,
closeness among the European countries seems to be stronger. What is
interesting, however, is the contraction of volume that characterizes the
reduced subspace presented in Figure 3. It shall be noticed that this is the
first 14-years interval that includes the failure of the Lehman Brothers and
the recent turmoil in the international banking sector.

\subsection{The Space Volume}

The observation of the plots presented in Figures 2 and 3 shows that a
generalized shortening of most of the distances leads to a contraction
effect in the volume of the \emph{3-}dimensional space. To capture the
extent of a contraction (or expansion) effect, we take the largest 3
eigenvalues that define the effective dimension of the market space and
compute its volume ($V$) as

\begin{equation}
\begin{array}{lll}
V_{t}=\sqrt[3]{\Pi_{i=1}^{3}\lambda _{t}(i)}\label{8.00} &  &
\end{array}%
\end{equation}

This measure $V$ is used throughout this paper as a reference for the
identification of the abnormal periods. As the plot in Figure 3 clearly
indicates, clustering emerges together with a contraction effect in the
shape of the reduced space. Such a contraction effect is registered as a
lower value of $V$. It so happens that, when the geometrical spaces are
built from cross-correlation based distances (Eq.1), expanded volumes
correspond to \emph{business as usual} periods while contractions have been
associated to the occurrence of economic crises (\cite{Araujo1},\cite%
{Araujo2}).

\begin{figure}[htb]
\begin{center}
\includegraphics[width=0.7\linewidth]{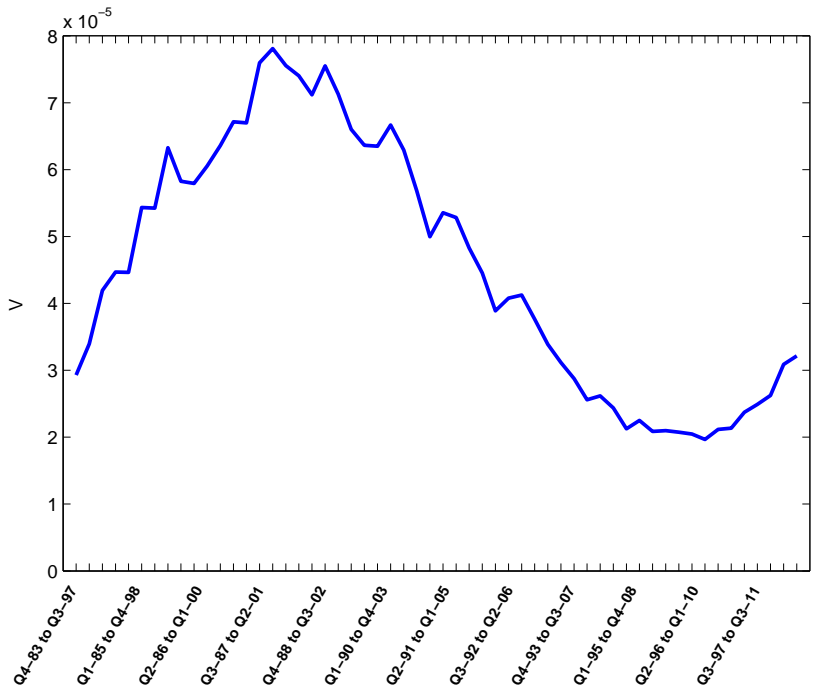}
\end{center}
\caption{The volume of the reduced \emph{3-}dimensional spaces along the
last 28 years}
\end{figure}

Figure 4 shows the behavior of $V$ along the 28 years. At each time interval
($t$), $V$ is computed from the distances measured over a moving window of
length 56 (14 years). The choice of such a wide window ensures statistical
robustness and allows for capturing the main differences in the behavior of $%
V$ when the latest nineties start to be taken into account.

The lowest values in the evolution of $V$ for the period under consideration
correspond to the 14-years slot Q2-1996 to Q1-2010. Figure 4 also shows that
the highest values of $V$ correspond to the time slot Q3-1987 to Q2-2001.
After that and up to 2010, one observes a persistent decreasing trend in the
values of $V$ and a contraction effect in the corresponding $\emph{3-}$%
dimensional spaces.

The evolution of $V$ confirms our previous results, identifying the major
changes in the 28-years period and highlighting the relevance of the recent
economic crises.

It shall be noticed that after 2009, the value of $V$ starts to increase,
meaning that a small recovery from the recent crises - translated into a
return to the situation of \emph{business as usual} - is taking place.

\section{The Topological Perspective}

In the previous section we have described the interbank market from a
geometrical approach. Here, departing from this geometrical setting, our
approach is driven by a topological perspective. Since the distances between
countries are properly defined distances, it gives a meaning to topological
tools in the study of the interbank market.

The first step in this direction concerns the definition of the
fully-connected and weighted networks of 24 reporting countries. These
networks are time-constrained since the strength of the links (their
weights) between each pair of countries depends on how distant they are in
the time-dependent \emph{3-}dimensional space.

\subsection{Networks of Bilateral Exposures}

Formally, countries $i$ and $j$ are linked by their \emph{bilateral exposure}
($B_{ij} $) computed as

\begin{equation}
B_{ij}(t)=B_{ji}(t)=\frac{1}{d^{(3)}_{ij}(t)}  \label{.7}
\end{equation}

where $d_{ij}^{(3)}(t)$ denotes the distance between countries $i$ and $j$
restricted to the \emph{3-}dimensional space and computed over a given time
interval ($t$). As usual, the inverse of the distance corresponds to the
intensity (strength) of a link between the involved entities. Therefore, the
bilateral exposure between each pair of countries, is given by the strength
of the link that binds the two countries, which depends on how close they
are in the $\emph{3-}$dimensional space.

\subsubsection{Spatial Dimension}

Taking the time interval of 110 quarters, from 1983 to 2011, we built the
network of 24 reporting countries connected by their bilateral exposures, as
the graphs in Figures 5 and 6 show. In the first graph, European countries
are represented in blue nodes and non-EU countries are colored red. There,
the size of a country ($i$) is given by its degree, i.e. by the total
strength of the links ($\sum_{j}^{N}B_{i(j)}$) in which this country ($i$)
participates. The width of the links are also proportional to their
strength, being the 10 stronger links of the whole network colored blue
while the remaining (weaker) links are colored cian. The graph in Figure 6
shows the corresponding network using the kamada-kawai layout \footnote{%
Kamada-kawai is a graph-drawing algorithm that after assigning forces among
the set of links and the set of nodes, pull them closer together or push
them further apart depending on the forces assigned to both nodes and links.}%

\begin{figure}[htb]
\begin{center}
\includegraphics[width=0.7\linewidth]{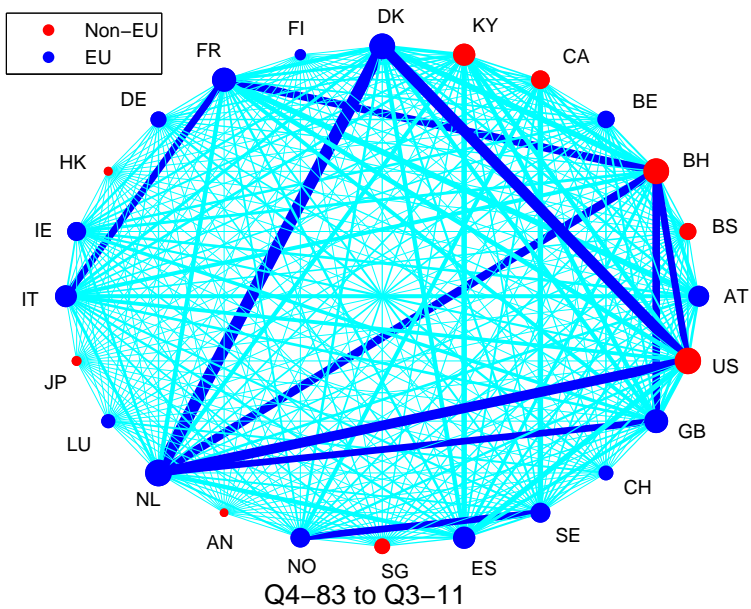}
\end{center}
\caption{The fully-connected network 1983-2011}
\end{figure}

These two graphs provide complementary information on the way the countries
are connected. The one in Figure 5 is aimed at emphasizing the heavily
connected countries and the intensity of country links. However, unless one
looks at the corresponding connections presented in the graph of Figure 6,
one is unable to identify the critical roles that some countries play in the
network structure. In this example, Figure 6 shows that when the 28-years
period is considered, the topological signature of this set of interbank
liabilities is characterized by the critical roles of Hong Kong (HK),
Finland (FI), Japan (JP) and Netherlands Antilles (AN).

\begin{figure}[htb]
\begin{center}
\includegraphics[width=0.8\linewidth]{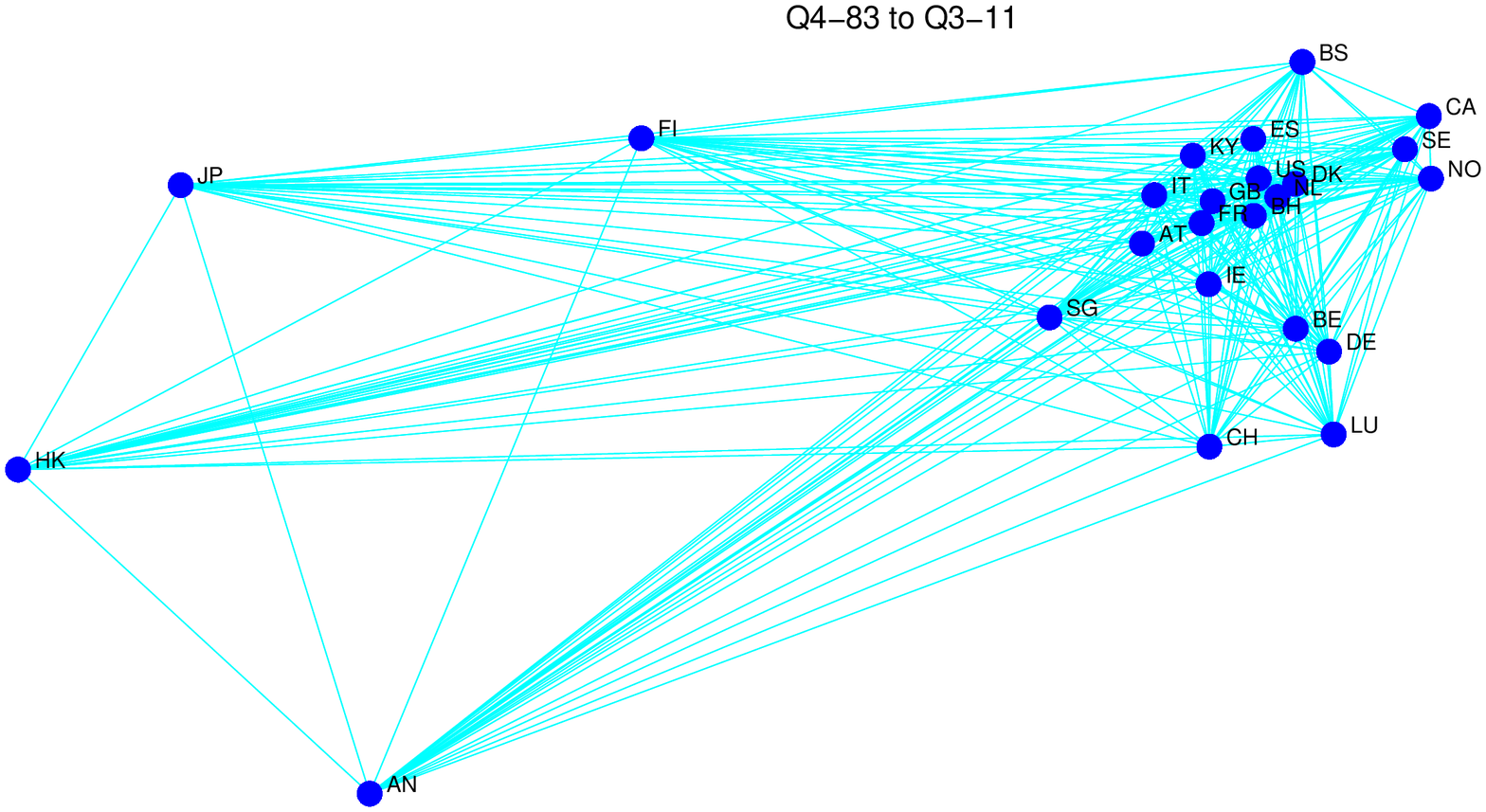}
\end{center}
\caption{The kamada-kaway network 1983-2011}
\end{figure}

Figure 5 informs that the strongest links are those involving nine
countries: United States (US), Denmark (DK), the Netherlands (NL), France
(FR), Italy (IT), Bahrain (BH) and the United Kingdom (GB). Norway (NO) and
Sweden (SE) are also among the strongly connected countries, although they
seem to form a separate group.

Albeit less evident in Figure 5, the Netherlands (NL), the United States
(US) and Bahrain (BH) hold the strongest four links. Despite the heavy
contributions of the United States (US) and Bahrain (BH), European countries
are, in general, those that display the strongest connections. Surprisingly,
this property does not hold for either Germany (DE) or Switzerland (CH).

\subsubsection{Temporal Dimension}

A further possibility in analyzing cross-country exposures is to consider
the time dimension of the data. The comparison of the data over time may
reveal to which countries bilateral exposure has become significantly larger
or smaller in specific time periods. Moreover, it might improve knowledge on
the impact of the recent financial crises to the interbank market structure
as a whole.

We proceed by developing, for specific time intervals of 56 quarters (14
years), the corresponding networks of country links. In so doing we are able
to observe the evolution of some structural patterns that emerge from the
dependencies between country debts over those time periods.

The observation of the graphs in Figure 7, where the periods 1983-1997 and
1997-2011 are represented, shows that the intensities of the links (the
degree of the nodes) have been greatly reinforced in the time period going
from 1997 to 2011, revealing that the amounts of liabilities between
countries became more strongly correlated during this last period.

\begin{figure}[htb]
\begin{center}
\includegraphics[width=0.45\linewidth]{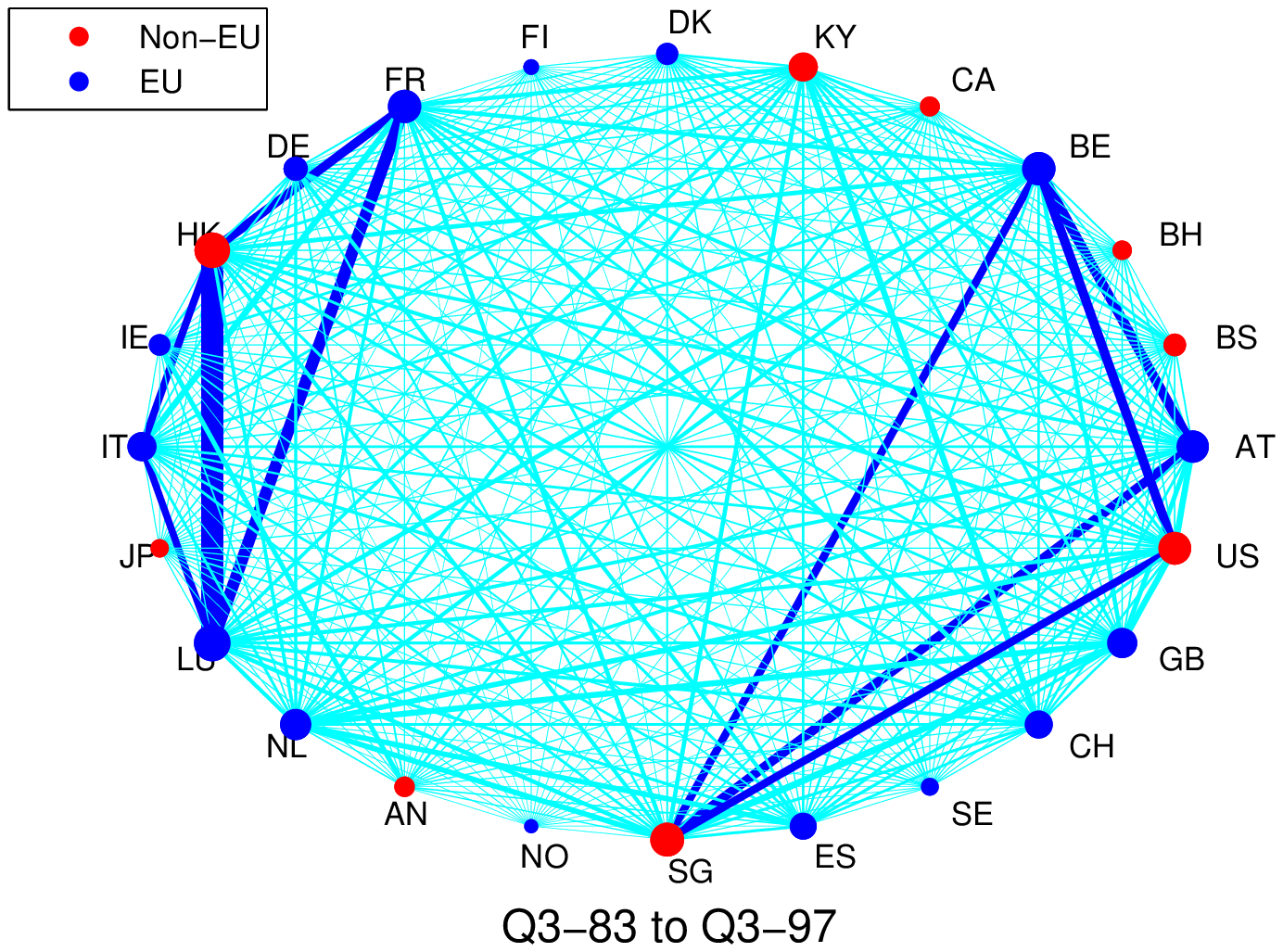}\includegraphics[width=0.45%
\linewidth]{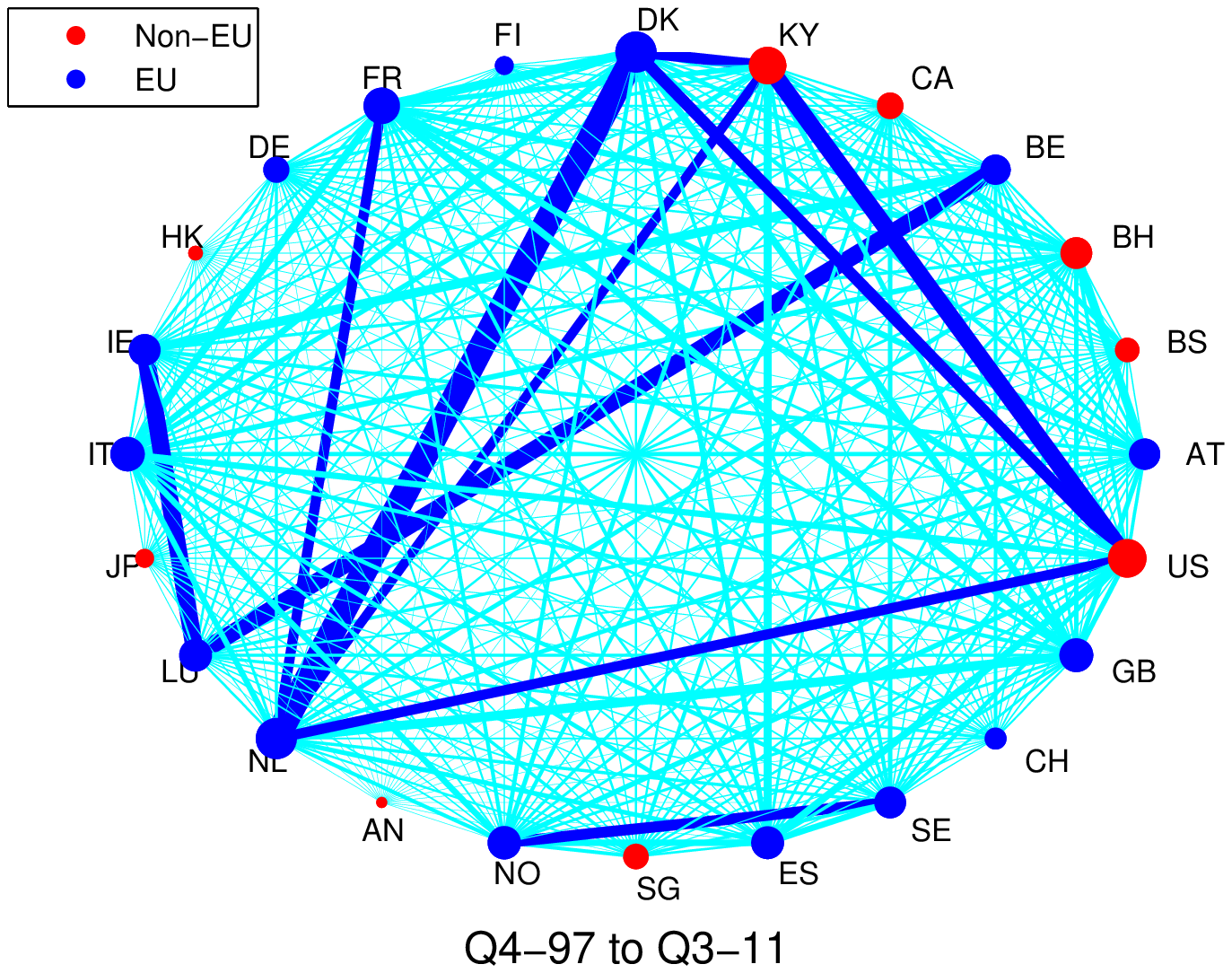}
\end{center}
\caption{The fully-connected networks 1983-1997 and 1997-2011}
\end{figure}

\begin{figure}[htb]
\begin{center}
\includegraphics[width=0.8\linewidth]{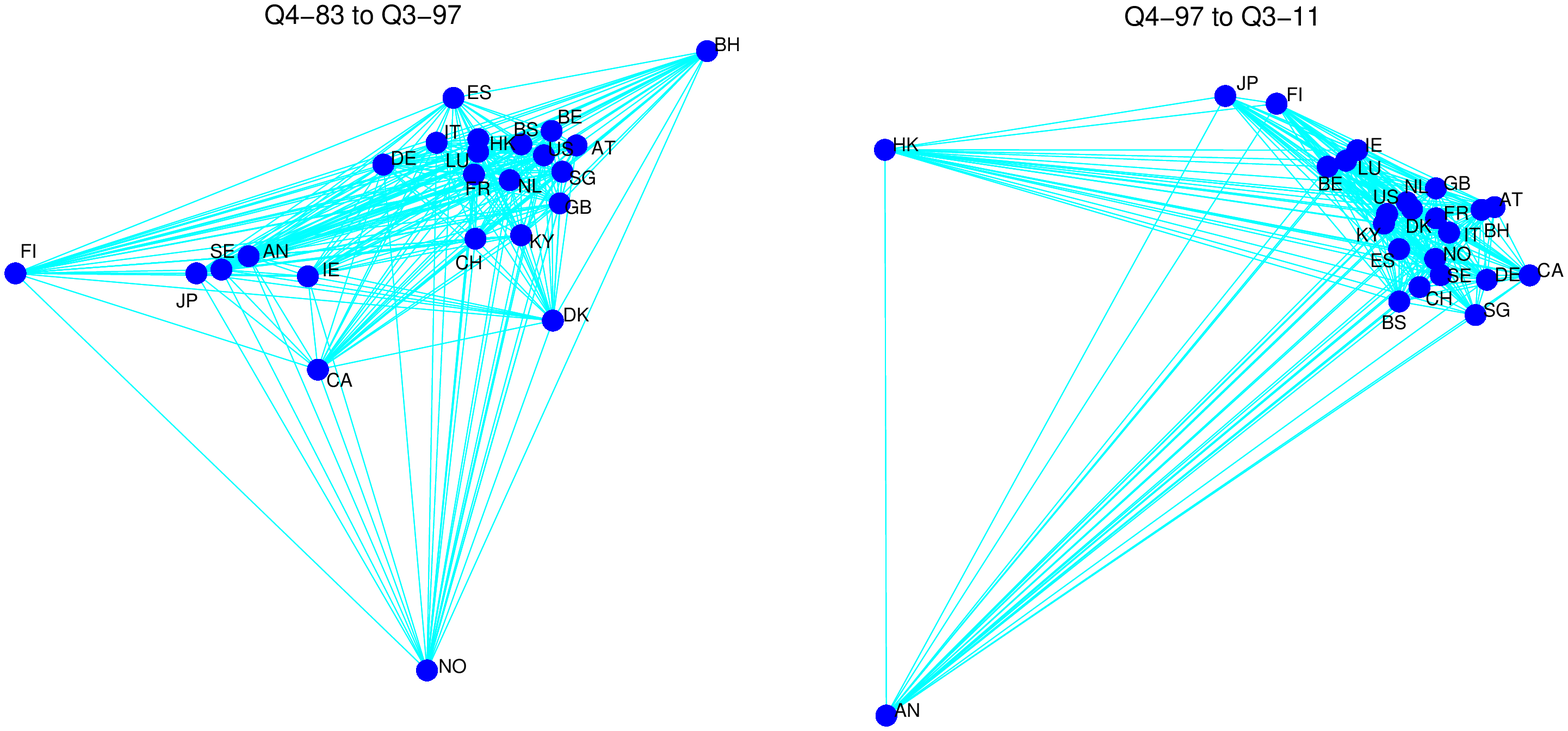}
\end{center}
\caption{The kamada-kaway networks 1983-1997 and 1997-2011}
\end{figure}

In the period 1983-1997 (l.h.s. in Figure 7) there are two separate sets of
strongly connected countries: France (FR), Hong Kong (HK), Luxembourg (LU)
and Italy (IT) in one side, and the United States (US), Belgium (BE),
Singapore (SG) and Austria (AT) in the other. The link between Hong Kong
(HK) and Luxemburg (LU) is the strongest one.

This situation undergoes a relevant change after 1997 (r.h.s. in Figure 7),
there being now three groups of strongly connected countries. The biggest
group comprises both EU and non EU countries: the United States (US),
Denmark (DK), the Netherlands (NL), France (FR) and the Caiman Islands (KY).
The second one is formed by Ireland (IE), Luxemburg (LU) and Belgium (BE).
There is also the geographically-oriented group, formed by Norway (NO) and
Sweden (SE). The Netherlands (NL) is the country reached by the stronger
links, including its remarkable connection to Denmark (DK).

When we look at the kamada-kawai graphs corresponding to each time period (Figure 8),
there is also a relevant change to be highlighted: the critical locations of
Norway (NO) and Finland (FI) are now replaced by the one of the Netherlands
Antilles (AN). Comparing the graphs in Figures 7 and 8, leads to the
following summarizing remarks:

\begin{itemize}
\item There is a generalized increase in the strength of the links (in the
network degree) in 1997-2011.

\item There is also a corresponding increase in clustering, in the sense
that the intensity of the links within the group of the strongly connected
countries are stronger than those that link them to any other (weakly
connected) country.

\item North EU countries (Norway and Finland) lost their critical positions,
being replaced by Hong Kong and the Netherlands Antilles.
\end{itemize}

To give a quantitative idea of the evolution of the structure of the network
of liabilities, we computed, for each country ($k$) and over the same time
intervals used in the last plots, its average degree ($\left\langle
B_{k}\right\rangle $). The results are presented in the histogram of Figure
9, where light colors correspond to periods close to 2011 and dark ones
indicate closeness to 1983.

The fact that the amounts of liabilities in the international interbank
market become high correlated in the last period gives place to a much
strongly connected network of countries. The histogram in Figure 9 confirms
that this effect is most apparent in North Europe countries such as Sweden
(SE), Finland (FI) Norway (NO) and Denmark (DK). The main feature in that
evolution relies on a kind of permutation taking place in the ranking of the
most connected countries. While the Netherlands Antilles (AN) and Hong Kong
(HK) suffer a relevant downgrade in that ranking, North EU countries moved
in the opposite direction, showing a remarkable increase in the degree of
connectivity.

\begin{figure}[htb]
\begin{center}
\includegraphics[width=0.7\linewidth]{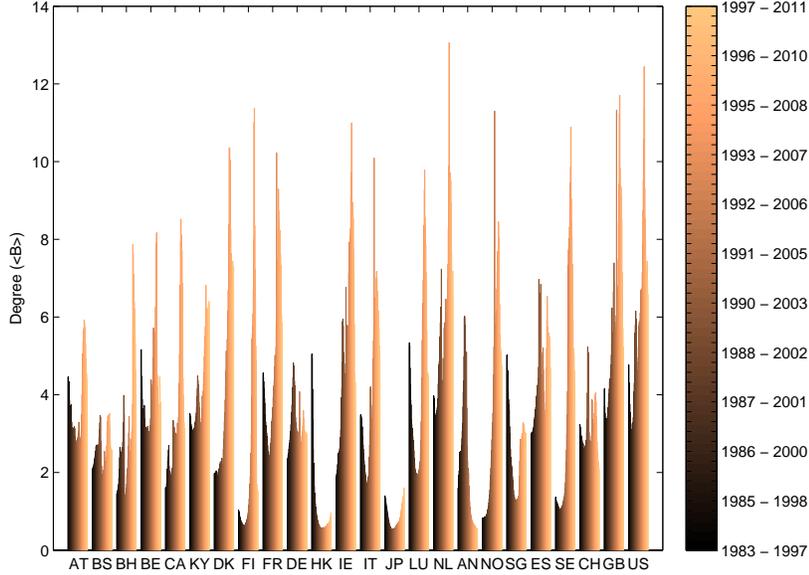}
\end{center}
\caption{The evolution of the degree of each country in the network}
\end{figure}

\subsection{Clustering and Closeness}

Although the graph structures presented so far provide pictorial and
detailed representations of the topological properties of the networks, to
have a wider view on their continuous evolution, there is a need to compute
- besides the degree of the nodes - the values of other topological
coefficients at each time period. These coefficients might also help to
figure out the extent to which the network structure was affected by the
recent financial crises. To this end, \emph{Closeness Centrality} and \emph{%
Continuous Clustering} were calculated over each 14-years period along the
110 time intervals from Q4-1883 to Q3-2011.

\subsubsection{Continuous Clustering}

Usually, topological coefficients apply to graph structures that are
connected and sparse. Using the distance matrix to construct the Minimal
Spanning Tree\footnote{%
A spanning tree of a connected and undirected graph is a subgraph that is a
tree and connects all the nodes together. The minimal spanning tree is
the spanning tree to which the sum of the weights of the links is minimal.}
connecting $N$ countries, as in Mantegna \cite{Mantegna}, we might start by
applying the graph theoretical notion of clustering to the spanning tree.
However this construction neglects part of the information contained in the
distance matrix. Instead we use the notion of \textit{Continuous Clustering }%
\cite{Vilela}.

Since $d^{(3)}_{ij}$ is the distance between the countries $i$ and $j$
restricted to the \emph{3-}dimensional space and $\overline{d^{(3)}}$ the
average distance, we define a function
\begin{equation}
V_{ij}=\exp \left( -\frac{d^{(3)}_{ij}}{\overline{d^{(3)}}}\right)
\label{2.12}
\end{equation}
which represents the \textit{neighbor degree} of the countries $i$ and $j$.
A continuous clustering coefficient is then defined by
\begin{equation}
C=\frac{1}{N(N-1)(N-2)}\sum_{i\neq j\neq k}V_{ij}V_{jk}V_{ik}  \label{2.13}
\end{equation}

The application of this measure to networks of stocks showed that a relevant
increase in the values of the coefficient of clustering is empirically
related to periods of market shocks or crises (\cite{Vilela},\cite{Araujo1},%
\cite{Araujo2}), capturing maximal information on market synchronization in
these periods and displaying a completely different behavior in normal
periods.

Figure 10 shows the continuous clustering coefficient calculated over a
moving window of 56 quarters. It displays an increasing trend with a steeper
slope starting in the time slot Q3-1992 to Q2-2006 and reaching the highest
value in 2008. After that, the values of $C$ indicate that the situation is
moving back to the one that characterizes the less synchronized behavior of
the earlier nineties.

The relative increase of $C$ reaches less than 10\% along the first 20 years
and after a stationary period that ends in 2006, $C$ shows a fast increase
of more than 15\% in just two years. After 2008, however, clustering - and
thus synchronous behavior in terms of correlated debts - starts to behave in
the opposite way, helping to shed some light on how the recent financial
crisis affected the interbank sector.

\begin{figure}[htb]
\begin{center}
\includegraphics[width=0.65\linewidth]{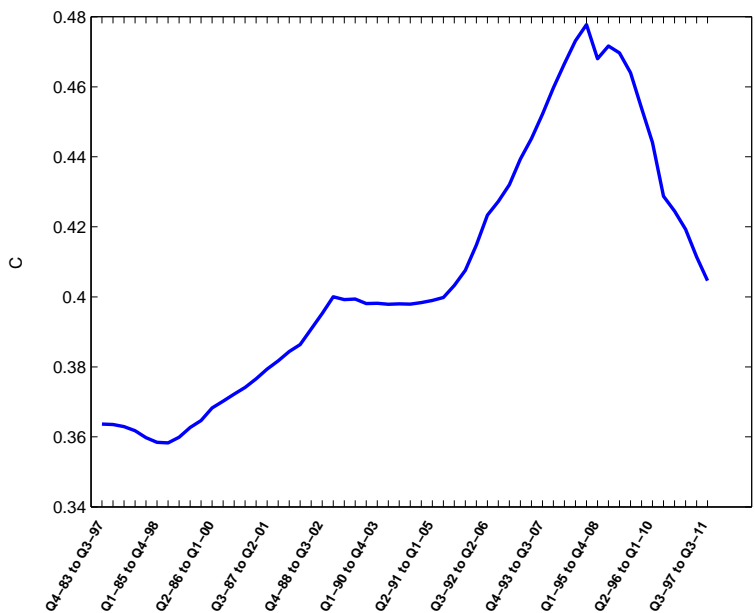}
\end{center}
\caption{The evolution of the coefficient of continuous clustering}
\end{figure}

Our results are in line with the work of Hattory \cite{Hattory}, who found
enough evidence of a significant augmentation of the values of the network
degree and a corresponding increase in the clustering coefficient over the
last decade\footnote{%
The author investigates the changes in cross-border bank exposures between
16 reporting countries over the period 1985-2006.}. Additionally, we
envision that such a clustering enlargement is the topological signature of
a period of mutation taking place in the market of cross-border liabilities
during the last 14 years.

\subsubsection{Closeness Centrality}

We have also computed the\emph{\ Closeness Centrality }of each country. In a
connected graph, the closeness centrality of a node $k$ is the mean geodesic
distance from $k$ to any other node, being given by

\begin{equation}
CC_{k}(t)=\sum_{j<>k}^{N}d_{k,j}^{(3)}(t)=\sum_{j<>k}^{N}\frac{1}{B_{k,j}(t)}
\end{equation}

where $d_{k,j}^{(3)}$ is the distance between $k$ and $j$ reduced to the $%
\emph{3-}$dimensional space. As a geodesic path is the shortest path between
a pair of nodes, closeness centrality is lower for nodes that are more
central in the sense of having a shorter average distance to the other nodes.

\begin{figure}[htb]
\begin{center}
\includegraphics[width=0.7\linewidth]{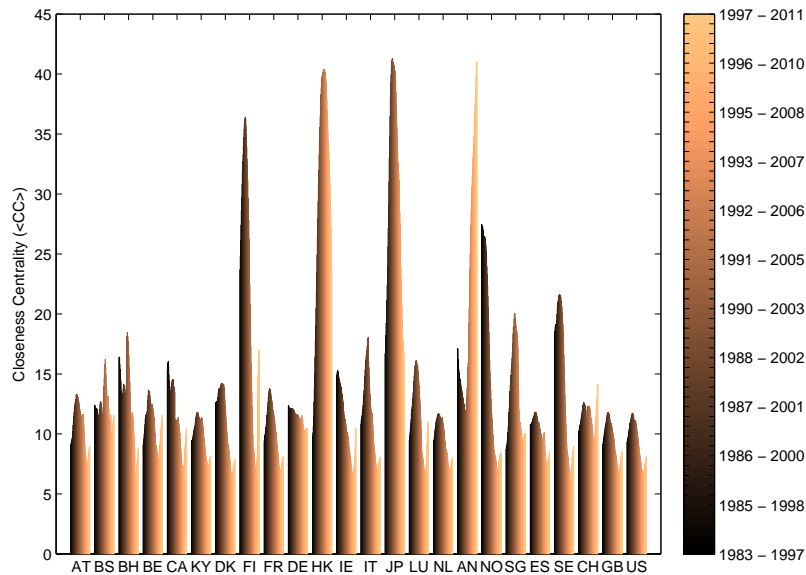}
\end{center}
\caption{The evolution of the closeness centrality of each country in the
network}
\end{figure}

The histogram presented in Figure 11 shows the values of closeness
centrality of each country along the whole 28 years. As usual, the values
were computed using a moving window of 56 quarters. Initially, both Finland
(FI) and Norway (NO) display high closeness. Later, they have been replaced
by Japan (JP) and Hong Kong (HK), and more recently, the country with the
highest closeness centrality turns to be the Netherlands Antilles (AN). This
result is in line with the evolution of the networks presented in Figure 8.
Noteworthy is the path followed by the most critical position in the
interbank network along the last 20 years, shifting from North Europe to an
off-shore country.

Such an important displacement could have been triggered by the European
Monetary System (EMS) crises in the early nineties, being then reinforced by
the Asian crisis in 1997. While EMS crisis was certainly associated to the
North Europe lost of closeness in the earlier nineties, the Japanese turmoil
in 1997-1998 might explain that closeness was shifted to an off-shore
country as the Netherlands Antilles.

Earlier we saw that Hong Kong (HK) and the Netherlands Antilles (AN) \
became the lowest connected countries and those with the most critical
positions (greatest closeness) in the network. According to the histogram in
Figure 11, such a critical role was earlier performed by the North European
countries Norway (NO) and Finland (FI).

In our example, the shift in the closeness centrality that was experienced
by the chain:
\begin{equation}
\textnormal{Norway} \Rightarrow \textnormal{Finland} \Rightarrow \textnormal{Japan} \Rightarrow \textnormal {HongKong} \Rightarrow \textnormal{Netherlands Antilles}
\end{equation}

reveals a moving trend grounded on successive financial crises and showing
part of their impact on the interbank international market.

Since the latest nineties, we know that crises were imposed either by
disarrangements of national markets or emerging economies like Russia and
Asian countries or by the performance of global players such as Japan. It is
also known that North European countries as Norway, Sweden and Finland
suffered important financial crises in the end of the eighties. As Vale
points out, the Norwegian crisis lasted from 1988 to 1993 and peaked in the
autumn of 1991 with the largest bank facing serious difficulties. It was
followed by the Finnish and Swedish crises, which significantly contributed
to the beginning of an important crisis in the European Monetary System
(EMS) \cite{Vale}.

A clarifying contribution comes from the work of Huizinga and co-workers,
who evaluated the extent to which the international banking flows come to be
reflected in tax policies of a large set of countries. In a recent paper
\cite{Huizinga}, they examine the relation between tax policies and the
amount of foreign liabilities in each national banking sector. As indicated,
foreign banks are expected to have relatively abundant opportunities to
shift profits when the host countries are characterized by low levels of
taxation. In this context, they found enough evidence on the inward profit
shifting role played by Norway and Hong Kong in the last decade.
Additionally, they recall that the most obvious recipients of inward profit
shifting are the low or no-tax off-shore financial centers as the Cayman
Islands and the Netherlands Antilles.

Here we envision that the network of cross-border liabilities channeled
through the banking system might be influenced by the sequence of financial
crises that impacted cross-border flows since the earlier nineties,
contributing to the emergence of the critical locations that correspond to
countries where tax policies provide opportunities to shift debts.

\section{Concluding Remarks}

In this paper we have investigated the modifications that may occur in the
structure of the cross-border financial linkages as a consequence of
economic crises, which in different historical periods, might be associated
to the opportunity to shift debts offered by countries with low tax policies.

Using time series of interbank liabilities conducted through the
international banking system, we have developed complete and weighted
networks of debt positions between 24 countries. These structures were
developed for \textit{(i)} a 28-years period that goes from 1983 to 2011 and
two 14-years periods \textit{(ii)} from 1983 to 1997, and \textit{\ (iii)}
from 1997 to 2011.

The geometrical analysis of \textit{(i)}, \textit{(ii)} and \textit{(iii)}
implied that most of the systematic information of the interbank
international market was contained in a $\emph{3-}$dimensional space, from
which a topological characterization could be conveniently carried out.

From the geometrical perspective, the reduced subspaces of countries showed
that the most eccentric positions are occupied by countries outside the
European Union. Moreover, it also showed that comparing \emph{(ii) }and
\emph{(iii)}, the main difference relies on the \emph{clustering trend }that
characterizes the later period, where the distances between countries seem
to be shortened, except for the few examples of the Netherlands Antilles and
Hong Kong. Such a clustering effect is followed by a contraction of the
volume ($V$) of the corresponding geometrical space. When $V$ is computed
over \emph{(i)}, its evolution confirms our previous results, identifying
the major changes in the 28-years period and highlighting the relevance of
the recent economic crises.

Our approach is complemented by a topological perspective. Since the
distances between countries are properly defined distances, it gives a
meaning to topological tools in the study of the interbank market. The first
step in this direction concerned the definition of fully-connected and
weighted networks of reporting countries. The analysis of the networks
obtained for \emph{(ii) }and \emph{(iii)} showed that: \emph{(a)} there is a
generalized increase in the network degree (strength of the links) in
1997-2011; \emph{(b)} there is also a corresponding increase in clustering,
and \emph{(c)} North EU countries lost their critical positions in the
network, being replaced by Hong Kong and the Netherlands Antilles.

Additionally, we argue that the chain of countries displaying the stronger
modifications in closeness and connectivity reflects the sequence of
financial crises that impacted cross-border flows since the earlier
nineties. We envision that the networks of cross-border liabilities channeled through the
banking system have critical locations that correspond to countries where
tax policies provide opportunities to shift debts.

As the data and the method suggest, such a feature is part of a mutation in
the structure of the cross-border interdependencies since the latest
nineties. Results highlight an important modification acting in the
financial linkages across countries in the period 1997-2011, and situate the
recent financial crises as a counterpart of a larger structural change going
on since 1997.

\bigskip

\textbf{Acknowledgement}: \emph{This work has benefited from partial
financial support from the Funda\c{c}\~{a}o  para a Ci\^{e}ncia e a Tecnologia-FCT, under
the 13 Multi-annual Funding Project of UECE, ISEG, Technical University of
Lisbon.}

\end{document}